\documentclass[useAMS,usenatbib]{mn2e}
\usepackage{graphicx,times}
\usepackage{amssymb}
\usepackage{amsmath}

\newcommand{\nh}{N_{\mathrm{H}}}
\newcommand{\chandra}{\textit{Chandra}}
\newcommand{\xmm}{\textit{XMM-Newton}}
\newcommand{\spitzer}{\textit{Spitzer}}
\newcommand{\herschel}{\textit{Herschel}}
\newcommand{\Lsun}{\rm{L_{\odot}}}

\title[Evolution of AGN: connecting X-ray and IR]{Evolution of luminosity function and obscuration of AGN: connecting X-ray and Infrared}
\author[Han et al.]
{Yunkun~Han$^{1,2,3}$\thanks{E-mail: hanyk@ynao.ac.cn},
Benzhong~Dai$^{4}$,
Bo~Wang$^{1,3}$,
Fenghui~Zhang$^{1,3}$,
Zhanwen~Han$^{1,3}$
\\
$^{1}$National Astronomical Observatories/Yunnan Observatory, the Chinese Academy of Sciences, Kunming 650011, China\\
$^{2}$Graduate University of Chinese Academy of Sciences, Beijing 100049, China\\
$^{3}$Key Laboratory for the Structure and Evolution of Celestial Objects, the Chinese Academy of Sciences, Kunming 650011, China\\
$^{4}$Department of Physics, Yunnan University, Kunming 650091, China
}

\begin{document}

\date{Accepted 2012 March 8.  Received 2012 March 6; in original form 2010 September 6}

\maketitle

\label{firstpage}

\begin{abstract}
We present a detailed comparison between the $2-10$ keV hard X-ray and infrared (IR) luminosity function (LF) of active galactic nuclei (AGN).
The composite X-ray to IR spectral energy distributions (SEDs) of AGN used for connecting the hard X-ray LF (HXLF) and IR LF (IRLF) are modeled with a simple but well tested torus model based on the radiative transfer and photoionization code CLOUDY.
Four observational determinations of the evolution of $2-10$ keV HXLF and six evolution models of the obscured type-2 AGN fraction ($f_2$) have been considered.
The $8.0$ and $15$~\micron~LFs for the total, unobscured type-1 and obscured type-2 AGN are predicted from the HXLFs, and then compared with the measurements currently available.
We find that the IRLFs predicted from HXLFs tend to underestimate the number of the most IR-luminous AGN.
This is independent of the choices of HXLF and $f_2$, and even more obvious for the HXLFs recently measured.
We show that the discrepancy between the HXLFs and IRLFs can be largely resolved when the anticorrelation between the UV to X-ray slope $\alpha_{\mathrm{ox}}$ and UV luminosity $L_{\rm UV}$ is appropriately considered.
We also discuss other possible explanations for the discrepancy, such as the missing population of Compton-thick AGN and possible contribution of star-formation in the host to the mid-IR.
Meanwhile, we find that the HXLFs and IRLFs of AGN can be more consistent with each other if the obscuration mechanisms of quasars and Seyferts are assumed to be different, corresponding to their different triggering and fueling mechanisms.
More accurate measurements of the IRLFs of AGN, especially that determined at smaller redshift bins and more accurately separated to that for type-1 and type-2, are very helpful for clarifying these interesting issues.
\end{abstract}


\section{INTRODUCTION}
\label{sect:intro}
Active galactic nucleus (AGN), compact regions at the center of active galaxies, releasing a great deal of energies in the form of radiation over the electromagnetic spectrum from radio, infrared, optical, ultraviolet, X-ray to $\gamma$-ray, are now believed to be powered by accretion of mass into the super-massive black holes (SMBHs).
In the local Universe, SMBHs are found to exist at the center of most massive galaxies.
There are good correlations between the mass of SMBHs and the properties of host galaxies \citep{Hopkins2007a,Kormendy2009a,Gultekin2009a,Merloni2010a}, such as the velocity dispersion \citep{Ferrarese00,Gebhardt00,Tremaine2002a}, mass \citep{Magorrian98,Haring2004a,Graham2004a}, or luminosity \citep{Kormendy95,Marconi03} of host bulge.
On the other hand, the AGN activity and star formation are found to peak at a similar redshift and decline towards low redshift simultaneously \citep{Hopkins04,Silverman08b,Aird10}.
Meanwhile, the mass density of local SMBHs in galaxy center is found to be consistent with that accreted by AGN throughout the history of the Universe \citep{Marconi04,Merloni04}. 
These correlations strongly support the idea that the growth of SMBHs should be coupled with the formation and evolution of galaxies \citep{Croton06,Bower06,DiMatteo05,DiMatteo07,Hopkins05b,Hopkins06d,Hopkins08}, although some authors \citep[e.g. ][]{Peng2007a,Jahnke2011a} argued a non-causal origin of them. 

While the important role of SMBHs, and so AGN, playing in the formation and evolution of galaxies has been well established, detailed mechanisms about this process are still largely unknown.
The luminosity functions (LF) of AGN, which describe the spacial density of AGN as a function of luminosity and redshift, is an important observable quantity for understanding the distribution and evolution of AGN.
It constrains the accretion history of SMBHs, and reveals the triggering and fueling mechanism of AGN and their co-evolution with host galaxies.
An observational determination of the bolometric LFs of AGN require multi-wavelength observations spanning the whole wavelength range of electromagnetic spectrum and sampling large comoving volume and luminosity range.
So, in practice, the LFs of AGN are measured independently from different wavelength bands such as radio \citep[e.g.][]{Nagar05}, infrared \citep[e.g.][]{Babbedge06,Brown06,Matute06}, optical \citep[e.g.][]{Fan01,Wolf03,Croom04,Richards06,Bongiorno07,Fontanot07,Shankar07}, soft X-ray \citep[e.g.][]{Miyaji00,Miyaji01,Silverman05b,Hasinger05}, hard X-ray \citep[e.g.][]{Ueda03,LaFranca05,Silverman05a,Silverman08b,Yencho09,Ebrero09,Aird10}, or emission lines \citep[e.g.][]{Hao05}.
However, due to the different selection effect suffered by different bands, the LFs of AGN measured from different bands are not necessarily consistent with each other.

Among various bands, X-ray, especially the hard X-ray band, is the most efficient for selecting AGN.
Recently, the evolution of the hard X-ray LF (HXLF) of AGN from $z\sim0$ to 5 is found to be best described by a luminosity dependent density evolution (LDDE) model.
According to this model, the spatial density of AGN with lower luminosity peaked at lower redshift than those with high luminosity, and the faint-end slope of the LFs is flattened as redshift increased \citep{Ueda03,Barger05,Hasinger05}.
This kind of so-called `cosmic downsizing' evolution trend of the AGN population has been further confirmed in radio and optical bands \citep{Cirasuolo05,Bongiorno07}.
These results revealed a dramatically different evolutionary model for Seyfert galaxies and quasars, and imply very different triggering, fueling and accretion mechanisms for the two classes of AGN.

Meanwhile, AGN are classified to two major classes according to their optical spectra.
Type-1 AGN exhibits both broad permitted lines and narrow forbidden lines in their spectra, while type-2 AGN presents only the narrow lines \citep{Khachikian74}.
\cite{Rowan-Robinson77} firstly put forward the idea that AGN are surrounded by dusty medium which absorbs their visible and ultraviolet light and then re-emits them in the mid-IR.
The extinction due to these obscuring medium is responsible for the distinction between type-1 and type-2 AGN.
Latterly, this idea was developed into the so-called unified model of AGN \citep{Pier92,Antonucci93,Maiolino95,Krolik99,Zhang06,Wang07}.
In the model, the differences between different types of AGN can be explained by the anisotropically distributed obscuring mediums (often visualized as a geometrically and optically thick torus comprised of dust and molecular gas) surrounding a basic black hole-accretion disk system, while different lines-of-sight into and through these obscuring mediums result in the diverse observational properties of AGN population.

However, the obscuration of AGN by anisotropically distributed gas/dust medium imply great systematic selection bias for understanding the properties and evolution of AGN.
Moreover, obscuring mediums around AGN are recently found to be distributed in a much more complex manner than a simple compact torus \citep{Risaliti02,Kuraszkiewic03,Risaliti05,Goulding09}, and may evolve with luminosity \citep{Ueda03,Steffen03,Hasinger04,Barger05,Simpson05} and redshift \citep{LaFranca05,Ballantyne06a,Treister06b,Hasinger08,Ebrero09}.
These results imply that the observational properties of AGN may vary significantly from object to object.
This complicates the understanding of the intrinsic characteristics of AGN and their correlations with the host galaxy.
On the other hand, current state-of-art synthesis models of cosmic X-ray background (CXRB) \citep[e.g.][]{Gilli07} show that a large population of heavily obscured Compton-thick AGN (with $\nh\le10^{24}\rm{cm}^{-2}$) are required to fit the CXRB spectrum. 
This population of Compton-thick AGN can be missed by even the deep hard X-ray surveys since they are deeply buried by obscuring medium.

Furthermore, recent results of \cite{Hasinger08} and \cite{Treister10} show that the fraction of absorbed AGN increases significantly with redshift to $z\sim2-3$, accompanied with the cosmic co-evolution of star-formation and AGN activity. 
These results support the idea that the obscuration of AGN cannot simply come from an unevolving torus employed by traditional unified model of AGN.
The obscuration mechanism of AGN with different triggering, fueling and accretion mechanisms may be different and associated with  their co-evolution with galaxies \citep{Davies06,Ballantyne06a,Ballantyne08}.

So, detailed studies of the obscuring medium around AGN, such as their geometry, distribution, composition, origin, and evolution, are very important \citep{Zhang04,Wang05,Liu11}.
The obscured or absorbed optical, ultraviolet, and X-ray radiation will be re-emitted in the infrared (IR).
IR bands represent an important complement for understanding the properties of obscuring medium around AGN and their co-evolution with host galaxies.
With the existing IR space telescope such as \spitzer, \herschel ~and forthcoming \textit{James Webb Space Telescope (JWST)}, our observation and understanding of AGN from the IR band will be largely improved.
Given the limitations suffered by X-ray observations, it is important to study the LFs and obscuration of AGN together and test the conclusions about their evolution, which are mainly based on observations in the X-ray band, in the IR band.

By using the spectral energy distributions (SEDs) modeled with a simple torus model, which is based on the radiative transfer and photoionization code CLOUDY, \cite{Ballantyne06b} can relate the X-ray and IR properties of AGN and explore the effects of parameters about obscuring medium.
They presented the mid-IR number counts and LFs for three evolution models of the  $f_2$ (equal to the covering factor under the unified model of AGN) that are constrained by the synthesis model of CXRB. 
The mid-IR number counts and LFs predicted from HXLF are in good agreement with direct IR observations, especially when assuming an inner radius ($R_{\mathrm{in}}$) of $10~\rm{pc}$ for the obscuring medium as expected if the obscuring material is connected to galactic-scale phenomenon. 
The mid-IR LFs of AGN are found to be a much better tool for determining the evolution of $f_2$ with $z$. 

\cite{Ballantyne06b} presented the mid-IR LFs for total AGN at different redshifts, but the observational mid-IR LFs of AGN \citep[i.e.][]{Brown06}, which is used to be compared with, are for type-1 AGN only.
After the work of \cite{Ballantyne06b}, some important improvements to the measurement of HXLF of AGN have been presented \cite[e.g.][]{Silverman08b,Ebrero09,Aird10}.
Furthermore, the actual evolution model of AGN obscuration is not necessarily within the three models proposed by \cite{Ballantyne06a}, other possibilities need to be tested for more reasonable conclusions.

In this paper, we present a more detailed comparison between the HXLFs and mid-IR LFs of AGN, which are connected by the composite X-ray to IR SEDs modeled with a modified version of the simple by well tested torus model of \cite{Ballantyne06b}. 
More observational determinations of the $2-10$ keV HXLF and the evolution models of $f_2$ have been considered.
The $8.0$ and $15$~\micron~ LF for the total, unobscured type-1 and obscured type-2 AGN are predicted from different combinations of HXLF and $f_2$, and then compared with current IR observational results.
Besides the measurement of \cite{Brown06}, the $15$~\micron~ LF given by \cite{Matute06} and recent results of \cite{Fu10} have been added for comparison.

We begin in Section \ref{sect:review} by reviewing current understanding of AGN evolution from X-ray band.
This include current observational determination of the evolution of the HXLF of AGN in Section \ref{ssect:hxlf}, and the evolution of AGN obscuration in Section \ref{ssect:f2}.
The detailed procedures of modeling the composite X-ray to IR SED of AGN, and our modifications to the original torus model of \cite{Ballantyne06b}, are presented in Section \ref{sect:seds}.
Section \ref{sect:connect} presents the method used to compute the IRLFs of type-1, type-2 and total AGN from different combinations of HXLF and $f_2$.
In Section \ref{sect:irlf_result}, we present our results and compare them with measurements from direct mid-IR observations to seek conclusions about the evolution of LFs and obscuration of AGN from combined views of hard X-ray and mid-IR.
We find that the mid-IR LFs predicted from HXLFs tend to underestimate the number of the most IR-luminous AGN, which is independent of the choices of HXLF and $f_2$, and even more obvious for the HXLFs recently determined.
In Section \ref{sect:discu}, we discuss explanations for this.
Finally, a summary of this paper is presented in Section \ref{sect:summary}.

Throughout this paper, we adopt a $H_0=70$~km~s$^{-1}$~Mpc$^{-1}$, $\Omega_{\Lambda}=0.7$, and $\Omega_{m}=0.3$ \citep{Spergel03} cosmology.
Minor differences in the cosmology have negligible effects on our conclusions.

\section{THE EVOLUTION OF AGN REVEALED FROM X-RAY BANDS}
\label{sect:review}
\subsection{The evolution of the HXLF of AGN}
\label{ssect:hxlf}
Strong X-ray emission is a unique indication of an AGN activity at the center of galaxies.
Deep X-ray surveys by \chandra~ and \xmm, which have already resolved most of the $2-10$ keV cosmic X-ray background (CXRB) into individual sources, found that most sources of CXRB are AGN.
X-ray, especially hard X-ray with energy $\ge$2 keV, are highly efficient for selecting AGN.
Both the moderately obscured ($\nh\le$ 10$^{23}$ \rm{cm}$^{-2}$) and low-luminosity sources commonly missed by optical observations can be selected from hard X-ray.
So, much more trustable evolution trends of the AGN can be revealed from hard X-ray observations.

As mentioned above, HXLF of AGN is found to be best described by a LDDE model.
However, the exact form of the evolution is still under debate, especially at high redshifts.
In this paper, we adopt the LDDE model given by \cite{Ueda03}, where the present-day HXLF is described as a smoothly-connected double power-law form:
\begin{equation}
\frac{{\rm d} \Phi (L_{\rm X}, z=0)}{{\rm d log} L_{\rm X}} 
= {\rm A} [(L_{\rm X}/L_{*})^{\gamma 1} + (L_{\rm X}/L_{*})^{\gamma 2}]^{-1},
\label{eq:hxlf0}
\end{equation}
where ${\gamma 1}$ is faint-end slope, ${\gamma 2}$ is the bright-end slope, $L_{*}$ is the characteristic break luminosity and ${\rm A}$ is a normalisation factor.
The evolution of LFs is given by
\begin{equation}
  \frac{{{\rm d}\Phi ({L_{\rm X}},z)}}{{{\rm d log} {L_{\rm X}}}} = \frac{{{\rm d}\Phi ({L_{\rm X}},0)}}{{{\rm d log} {L_{\rm X}}}}e(z,{L_{\rm X}}),
\label{eq:hxlfz_LDDE}
\end{equation}
where the evolution term is given by
\begin{equation}
e(z,{L_{\rm X}}) = \left\{ {\begin{array}{*{20}{l}}
   {{{(1 + z)}^{p1}}} & {[z < {z_{\rm c}}({L_{\rm X}})]}  \\
   \\
   {e({z_{\rm c}}){{[\frac{{1 + z}}{{1 + {z_{\rm c}}({L_{\rm X}})}}]}^{p2}}} & {[z \ge {z_{\rm c}}({L_{\rm X}})]}.  \\
\end{array}} \right.
\label{eq:ez_LDDE}
\end{equation}
The cutoff redshift $z_{\rm c}$, with a dependence on the luminosity starting from a characteristic luminosity ${L_a}$, are given by a power law of $L_{\rm X}$:
\begin{equation}
{z_{\rm c}}({L_{\rm X}}) = \left\{ {\begin{array}{*{20}{l}}
   {z_{\rm c}^*} & {\left( {{L_{\rm X}} \ge {L_a}} \right)}  \\
   \\
   {z_{\rm c}^*{{\left( {\frac{{{L_{\rm X}}}}{{{L_a}}}} \right)}^\alpha }} & {\left( {{L_{\rm X}} < {L_a}} \right)},  \\
\end{array}} \right.
\label{eq:zc_LDDE}
\end{equation}
where $\alpha$ measures the strength of the dependence of $z_{\rm c}$ with luminosity.

\begin{figure}
  \begin{center}
    \includegraphics{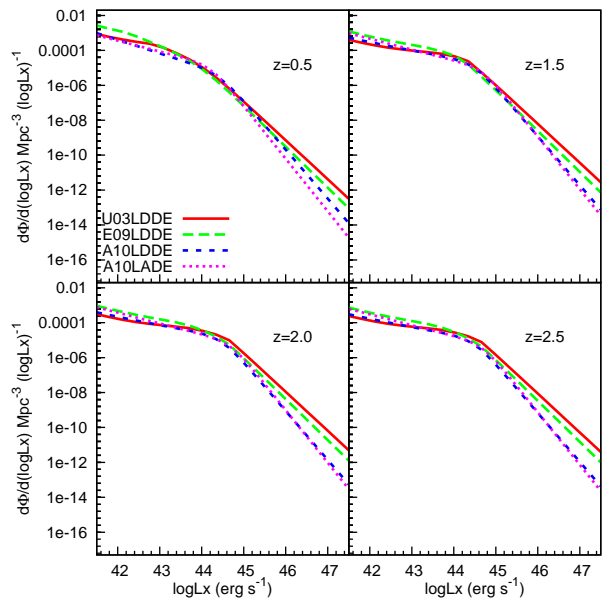}
  \end{center}
  \caption{HXLF of AGN at $z=0.5, 1.5, 2.0$ and $2.5$ as given by the LDDE model of \citet{Ueda03} (red solid line, noted as U03LDDE), \citet{Ebrero09} (green dashed line, noted as E09LDDE), \citet{Aird10} (blue short dashed line, noted as A10LDDE) and the LADE model of \citet{Aird10} (purple dot line, noted as A10LADE), respectively.}
  \label{fig:hxlf}
\end{figure}

Recently, \cite{Ebrero09} re-measured the HXLF of AGN by using the \xmm~Medium Survey \cite[XMS,][]{Barcons07} and other highly complete deeper and shallower surveys to assemble an overall sample of $\sim450$ identified AGN in the $2-10$ keV band, which is one of the largest and most complete sample up to date.
\cite{Aird10} presented a new observational determination of the evolution of the $2-10$ keV HXLF of AGN by using data from many surveys including 2 Ms \chandra\ Deep Fields and the AEGIS-X 200 ks survey.
These, combined with a sophisticated Bayesian methodology, allow them to do a more accurate measurement of the evolution of the faint end of the HXLF.
They found that the evolution of HXLF are best described by a so called luminosity and density evolution (LADE) model, rather than the LDDE model.
The LADE model is a modified Pure luminosity evolution (PLE) model.
According to the PLE model \citep{Ueda03}, the evolution of HXLF with redshift are described by allowing the characteristic break luminosity $L_{*}$ in the present-day HXLF as given by Eq. (\ref{eq:hxlf0}) to evolve as
\begin{equation}
\log L_*(z) = \log L_0 - \log\left[ \left(\frac{1+z_c}{1+z}\right)^{p_1} + \left(\frac{1+z_c}{1+z}\right)^{p_2} \right]
\label{eq:Lstar_PLE}
\end{equation}
where the parameter $z_c$ controls the transition from the strong low-$z$ evolution to the high-$z$ form. 
The LADE model are constructed by additionally allowing for overall decreasing density evolution with redshift, i.e. allowing the normalization constant $A$ in the present-day HXLF as given by Eq. (\ref{eq:hxlf0}) to evolve as
\begin{equation}
  \log A(z)= \log {\rm A_0} + {\rm d}(1+z)
\end{equation}
where ${\rm d}$ is an additional parameter describing the overall density decreasing. 

In Fig. \ref{fig:hxlf}, we show the  HXLF of AGN at $z=0.5, 1.5, 2.0$ and $2.5$ as given by the LDDE model of \cite{Ueda03}, \cite{Ebrero09}, \cite{Aird10}, and the LADE model of \cite{Aird10}.
The LADE modeling of the HXLF retains the same shape at all redshifts, but undergoes strong luminosity evolution out to $z\sim$ 1.
Meanwhile, the HXLF undergoes overall negative density evolution with increasing redshift.
Fig. \ref{fig:hxlf} clearly shows that the HXLF of \cite{Aird10} is different from those of the others at high luminosity.

Different HXLFs have very different implications for the AGN populations, such as their lifetimes, duty cycles, fueling, triggering and evolution.
Further complemental views from other wavelength bands, such as IR, are important for a fully understanding of the evolution of AGN populations.
In this paper, the four modelings of HLXF mentioned above have been used to predict the evolution of mid-IR LFs of type-1 and/or type-2 AGN, respectively.

\subsection{The evolution of the obscuration of AGN}
\label{ssect:f2}
Since there is a good correspondence between the AGN with $\nh\ge$ 10$^{22}$ \rm{cm}$^{-2}$ and those optically identified as being of type-2 \citep{Tozzi06}, type-2 AGN is commonly defined as those with absorbing column densities $\nh\geq 10^{22}$~\rm{cm}$^{-2}$ in the X-ray band.
According to the unified model of AGN, $f_2$ approximately equal to the covering factor of the gas with $\nh\geq 10^{22}$~\rm{cm}$^{-2}$ around the AGN.

\begin{figure*}
  \begin{center}
    \includegraphics[scale=1.2]{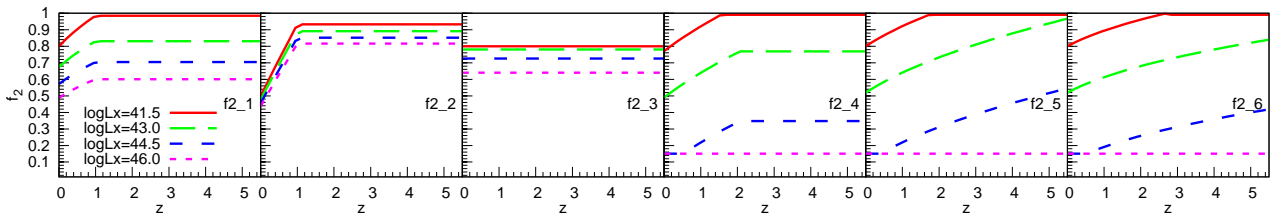}
  \end{center}
  \caption{The six evolution models of $f_2({\rm log}L_{\rm X},z)$ used in this paper are shown at ${\rm log}L_{\rm X}$=41.5 (red solid line), 43.0 (green dashed line), 44.5 (blue short dashed line) and 46.0 (purple dot line), respectively. The first three, which are constrained by CXRB spectrum and X-ray number count as given by \citet{Ballantyne06a}, are noted as `f2\_1', `f2\_2', and `f2\_3', respectively. The last three, which are constructed according to recent measurement of \citet{Hasinger08}, are noted as `f2\_4', `f2\_5', and `f2\_6', respectively.(see text for more detailed explanations for these evolution models.)
}
  \label{fig:f2}
\end{figure*}

By using a population synthesis model of CXRB, \cite{Ballantyne06a} constrained the evolution of $f_2$ as a function of both $z$ and $L_{\rm X}$.
They presented three parameterizations for the evolution of $f_2({\rm log}L_{X}, z)$ that could fit the observed shape of the CXRB and X-ray number counts of AGN.
The first one (shown in Fig. \ref{fig:f2}, and noted as `f2\_1'), with a moderate redshift evolution, is given as:
\begin{equation}
  f_2 = {\rm K_1} (1+z)^{0.3} ({\rm log} L_{\rm X})^{-4.8},
\label{eq:f2_1}
\end{equation}
where ${\rm K_1}$ is a constant defined by $f_2({\rm log}L_{\rm X}=41.5,z=0)=0.8$, which is based on observations in the local Universe.
The second one (shown in Fig. \ref{fig:f2}, and noted as `f2\_2') with a more rapid redshift evolution, is given as:
\begin{equation}
  f_2 = {\rm K_2} (1+z)^{0.9} ({\rm log} L_{\rm X})^{-1.3},
\label{eq:f2_2}
\end{equation}
where ${\rm K_2}$ is based on the Sloan Digital Sky Survey (SDSS) measurement of $f_2({\rm log}L_{\rm X}=41.5,z=0)=0.5$ by \cite{Hao05}.
In the above two cases, the $z$ evolution is halted at $z=1$, because there is no constraint on $f_2$ at higher redshifts.
The last one (shown in Fig. \ref{fig:f2}, and noted as `f2\_3'), which is considered as a null-hypothesis, assumes that $f_2$ does not evolve with redshift, and given as:
\begin{equation}
  f_2={\rm K_3} \cos^2\left(\frac{{\rm log} L_{\rm X} - 41.5}{9.7}\right),
\label{eq:f2_3}
\end{equation}
where ${\rm K_3}$ is determined by $f_2({\rm log}L_{\rm X}=41.5,z=0)=0.8$.

The fraction of type-2 AGN can also be measured directly from observations in different bands.
For example, in the optical, AGN can be selected using their high ionization lines to construct the standard diagnostic diagrams \citep{Baldwin81,Kewley01,Kauffmann03c,Kewley06}.
Furthermore, the ratio of narrow-line and broad-line AGN can be measured as a function of $z$ and the luminosity of emission lines (such as [O III] $5007 \text{\AA}$ line), which can be used as AGN power indicators.
However, significant limitations of optical selection and classification of AGN have been noticed \citep[e.g.][]{Moran02,Netzer06,Rigby06}.
The nuclear emission can be obscured by the torus and/or outshone by the host-galaxy light.
Alternatively, AGN can be efficiently selected in the X-ray, and the X-ray luminous AGN can be classified to absorbed and unabsorbed according to their absorbing column densities ${\rm log} \nh<22$ or $>22$.
X-ray selection of AGN is suffered by the limited sensitivity of telescopes, which are only sensitive at $\leq10$ keV.
So, a significant fraction of absorbed objects, especially the large number of Compton-thick AGN with ${\rm log} \nh>24$ predicted by the population synthesis model of CXRB, may be missed by current hard X-ray selection of AGN.

The combination of X-ray and optical criteria is a much more robust method for the selection and classification of AGN.
Recently, \cite{Hasinger08} have presented a new determination of the fraction of absorbed sources as a function of X-ray luminosity and redshift from a sample of 1290 AGN.
They are selected in the $2-10$ keV band from different flux-limited surveys with very high optical identification completeness, and grouped into type-1 and type-2 according to their optical spectroscopic classification and X-ray absorption properties.
So, the evolution of AGN absorption with luminosity and redshift is determined with higher statistical accuracy and smaller systematic errors than previous results.
The absorbed fraction is found to decrease strongly with X-ray luminosity, and can be represented by an almost linear decrease with a slope of $0.281\pm0.016$.
Meanwhile, it increase significantly with redshift as $\sim(1+z)^{0.62\pm0.11}$ from $z=0$ to $z\sim2$.
On the other hand, the evolution of the absorbed AGN fraction over the whole redshift from  $z=0$ to $z\sim5$ can also be described as $\sim(1+z)^{0.48\pm0.08}$, or $\sim(1+z)^{0.38\pm0.09}$ when data with crude redshifts are excluded.

These findings may have important consequences for the broader context of AGN and galaxy co-evolution. 
According to the results of \cite{Hasinger08}, we have constructed three new evolution models of AGN obscuration, which are expressed as
\begin{equation}
f_2=-0.281({\rm log}L_{\rm X}-43.75)+0.279(1+z)^{0.62},
\label{eq:f2_4}
\end{equation}
\begin{equation}
f_2=-0.281({\rm log}L_{\rm X}-43.75)+0.308(1+z)^{0.48},
\label{eq:f2_5}
\end{equation}
\begin{equation}
f_2=-0.281({\rm log}L_{\rm X}-43.75)+0.309(1+z)^{0.38},
\label{eq:f2_6}
\end{equation}
and noted as `f2\_4', `f2\_5', and `f2\_6', respectively (shown in Fig. \ref{fig:f2}).
Due to the simple linear dependence on luminosity, the type-2 AGN fraction will quickly become zero as luminosity increasing.
According to recent results of \cite{Brusa10}, the fraction of the obscured AGN population at the highest ($L_{\rm X}>10^{44} {\rm erg}~{\rm s}^{-1}$) X-ray luminosity is $\sim15\%-30\%$.
So, we have set a lower limit of $0.15$ for the evolution of $f_2({\rm log}L_{\rm X},z)$ to stand for a flattening of the decline at the highest luminosities as expected.
Naturally, an upper limit of $1$ is forced at all cases.

Except for $f_2({\rm log}L_{\rm X},z)$, additional assumptions are needed to determine a specific distribution of $\nh$.
The exact distribution of $\nh$ is unknown except for local bright Seyfert 2s \citep[e.g.][]{Risaliti99}, but the covering factor is a useful parameter for its theoretical description.
Here, we use a simple assumption about the distribution of $\nh$ following \cite{Ballantyne06b}.
In the `simple $\nh$ distribution', ten values of $\nh$ are considered: ${\rm log}(\nh/{\rm cm}^{-2})=20,20.5,\ldots,24.0,24.5$, and a type-1 AGN is assumed to have an equal probability $p$ of being absorbed by columns with ${\rm log} \nh < 22$.
Likewise, a type-2 AGN has an equal chance of being absorbed by columns with ${\rm log} \nh \ge 22$:
\begin{equation}
{\rm log}(\nh/{\rm cm}^{-2}) = \left\{ {\begin{array}{*{20}{l}}
  {20.0, \ldots ,21.5},&{p = \frac{{1 - {f_2}({\rm{log}}{L_{\rm{X}}},z)}}{{4.0}}} \\ 
  \\
  {22.0, \ldots ,24.5},&{p = \frac{{{f_2}({\rm{log}}{L_{\rm{X}}},z)}}{{6.0}}} 
\end{array}} \right.
\label{eq:simple}
\end{equation}
Since $f_2({\rm log}L_{\rm X},z)$ depends on ${\rm log}L_{\rm X}$ and $z$, the distribution of obscuring medium around AGN evolves with both ${\rm log}L_{\rm X}$ and $z$.
It is worth noting that this simple assumption about the distribution obscuring medium with different $\nh$ is only used to construct $\nh$-averaged SEDs (described in the next Section).
We do not expect it to give a correct fraction of Compton-thick AGN.
In fact, there are AGN with estimated column ${\rm log} \nh \ge 25$ \cite[e.g. NGC 1068;][]{Matt97}.
The inclusion of very Compton-thick obscuring medium dramatically increase the computation time of SEDs.
However, they have important effects mainly in the far-IR, but only ignorable effects in the mid-IR which we are mostly interested in currently.

\section{MODELLING THE SPECTRAL ENERGY DISTRIBUTION OF AGN}
\label{sect:seds}
To predict the IR properties of AGN, we must know the relation between IR and X-ray luminosity of AGN.
This can be given by the SEDs of AGN with different X-ray luminosities.
The SEDs of AGN can be obtained from observations or theoretical calculations.
The observational SEDs have the advantage of being based on observations of real AGN.
However, the number of observed objects is limited, and they only cover a narrow range of luminosity, redshift and wavelength.
Alternatively, we can use theoretical dusty torus emission models, which include a detailed radiative transfer calculation, to compute the expected IR SED for a given X-ray luminosity.
However, most radiative transfer calculation for IR dust emission of AGN do not include detailed considerations of gas and its interaction with dust \cite[e.g.][]{Treister04,Treister06a,Nenkova08a,Nenkova08b}.
The gas and dust are expected to be interacting with each other, and gas is responsible for the absorption of X-ray.
So, for a reasonable connection of the X-ray and IR properties of AGN, gas and their interaction with dust must be considered.

Here, the calculation of AGN SEDs is performed by using the photoionization code CLOUDY v.\ 07.02.01 \citep{Ferland98}, following a procedure similar to that of \citet{Ballantyne06b}, but with some simplifications.
In CLOUDY, the atomic gas physics along with the detailed dust radiation physics, such as polycyclic aromatic hydrocarbon (PAH) emission and emission from very small grains, have been self-consistently considered.
In addition, many important physical properties of the obscuring medium around AGN, such as its distance from the central engine, gas density, distribution and gas/dust ratio, can be varied freely, and so explored extensively.
CLOUDY is a one-dimensional radiative transfer code, and the methods we employed to model the SEDs of AGN are less sophisticated than those used by \cite{Treister06a} or \cite{Nenkova08a}.
However, \citet{Ballantyne06b} showed that the SEDs, when averaged over a $\nh$ distribution, have very similar properties to the ensemble of AGN found in the deep surveys of \chandra, \xmm~ and \spitzer.

\subsection{Construction of CLOUDY model}
\label{sssect:cloudy}
To construct a CLOUDY model, three ingredients must be specified.
Firstly, the shape and intensity of the radiation source, which define the incident continuum, must be set.
The intrinsic spectrum of AGN is described by a multi-component continuum typical for AGN, which extend from 100 keV to $>1000$~\micron.
Specifically, the `Big Bump' component, peaking at ≈1 $\rm{Ryd}$, is a rising power law with a high-energy exponential cutoff and parameterized by the temperature of the bump.
The big blue bump temperature is set to be a typical value of $10^5$K.
The X-ray to UV ratio $\alpha_{\mathrm{ox}}$, which is defined by 
\begin{equation}
  \alpha_{\mathrm{ox}}=\frac{{\rm log}({L_{2{\rm\,keV}}/L_{2500\rm{\mathring{A}}}})}{{\rm log}({\nu_{2{\rm\,keV}}/\nu_{2500\rm{\mathring{A}}}})}=0.3838~{\rm log}\left({\frac{L_{2{\rm\,keV}}}{L_{2500\rm{\mathring{A}}}}}\right),
\end{equation}
have an important effect on the resulting X-ray to IR ratio.
Especially, there are evidences \cite[e.g.][]{Steffen06,Hopkins07a,Vagnetti10} that this parameter may be anticorrelated with the UV luminosity of AGN.
To explore the effects of this important parameter, we set $\alpha_{\mathrm{ox}}$ to be a constant value of $-1.5$, $-1.4$ and $-1.3$, respectively.
We also tested the $\alpha_{\mathrm{ox}}-L_{\rm UV}$ relation presented in \cite{Hopkins07a}, which is given by
\begin{equation}
\alpha_{\rm ox}=-0.107\,\log(\frac{L_{2500\rm{\mathring{A}}}}{{\rm erg\,s^{-1}\,Hz^{-1}}})+1.739,
\label{eq:alphaox}
\end{equation}
and determined specifically for unobscured (type-1) quasars.
This results in a luminosity-dependent shape of the input SED from AGN center.
The low-energy slope of the Big Bump continuum $\alpha_{\mathrm{uv}}$ is set to be the default value of $-0.5$.
The X-ray photon index is assumed to be $\Gamma=1.9$, and so the energy index ${\alpha _\mathrm{x}}=1-\Gamma=-0.9$.
The full continuum is the sum of two components as given by
\begin{equation}
  {f_\nu } = {\nu ^{{\alpha _{\mathrm{uv}}}}}\exp ( - h\nu /K{T_{\mathrm{BB}}})\exp ( - K{T_{\mathrm{IR}}}/h\nu ) + a{\nu ^{{\alpha _\mathrm{x}}}},
\end{equation}
where ${T_{\mathrm{BB}}}$ is the temperature of Big Bump and the coefficient $a$ is adjusted to produce the correct $\alpha_{\mathrm{ox}}$ for the case where the Big Bump does not contribute to the emission at 2 keV.
The Big Bump component is assumed to have an IR exponential cutoff at $KT_{\mathrm{IR}} = 0.01 \rm{Ryd}$ ($1\rm{Ryd}\sim13.6~eV$).\footnote{See CLOUDY document for more detailed explanations for the construction of this AGN spectrum.}
Finally, this spectrum is scaled to have a luminosity of $L_{2500\rm{\mathring{A}}} (\rm erg\;s^{-1}\;Hz^{-1})$.

The second ingredient of a CLOUDY model is the chemical composition of the obscuring medium.
A gaseous element abundance similar to that of Orion Nebula is assumed.
The size distributions and abundances of graphitic, silicate and PAHs grains are also set to be similar to that of Orion Nebula.
The obscuring medium is assumed to distribute uniformly and have a constant hydrogen density $n_{\rm{H}}$ of $10^{4} \rm{cm}^{-3}$.

The last ingredient of a CLOUDY model is the geometry of the obscuring medium.
Here, the obscuring medium is assumed to be $R_{\mathrm{in}}~\rm{pc}$ away from the center and with a column density of $\nh$.
On the other hand, to be consistent with the unified model, \cite{Ballantyne06b} have set the covering factor of the obscuring medium to $f_2$ when $\nh \geq 10^{22}$~\rm{cm}$^{-2}$ or $1-f_2$ otherwise, in CLOUDY model.
Since $f_2$ depends on both luminosity and redshift, the CLOUDY simulation needs to be done for each luminosity and redshift, respectively.
This would result in a great number of CLOUDY models.
However, in CLOUDY models the covering factor only has second-order effects on the spectrum through changes in the transport of the diffuse emission.
So, we just use the default geometric covering factor of unity (the shell fully covers the continuum source) but a radiative covering factor of zero, i.e. an open geometry is assumed, and the reflected radiation can be obtained as well.
The effects of covering factor on the diffuse and reflected emissions are considered after the CLOUDY simulation as described int the next section.

\subsection{CLOUDY model grids and construction of AGN SEDs}
\label{sssect:grids}
The CLOUDY models are built for $L_{2500\rm{\mathring{A}}} (\rm erg\;s^{-1}\;Hz^{-1})$ \footnote{To explore the effects of $L_{\rm UV}$-dependent $R_{\mathrm{in}}$ and $\alpha_{\mathrm{ox}}$ on the resulting SEDs, $L_{2500\rm{\mathring{A}}}$ instead of $L_{\rm X}$ is used to define the luminosity of input SED.} from 27 to 34 (in steps of 0.25), and ${\rm log} (\nh/{\rm cm}^{-2})$ from 20.0 to 24 (in steps of 0.5).
Following \cite{Ballantyne06b}, we firstly set $R_{\mathrm{in}}$ to be $10~\rm{pc}$.
However, we found that the temperatures of grains will be much higher than their sublimation temperatures at the high-luminosity end if a constant $R_{\mathrm{in}}$ of $10~\rm{pc}$ is assumed.
We have practically found a luminosity-dependent $R_{\mathrm{in}}$, which is given by
\begin{equation}
  R_{\mathrm{in}}=10*[\frac{\nu L_{\nu}(2500\rm{\mathring{A}})}{10^{46}}]^{1/2} {\rm pc},
\label{eq:Rin}
\end{equation}
to fix this problem.
The CLOUDY models are also built by assuming luminosity-dependent $R_{\mathrm{in}}$ for comparison.
Finally, as mentioned above, the CLOUDY models are built for four choices of $\alpha_{\mathrm{ox}}$, respectively.

For each CLOUDY model, three kinds of SEDs are predicted: the attenuated incident continuum, diffuse continuum and reflected continuum.
The SEDs with different $\nh$ correspond to observations from different direction.
However, according to the unified model of AGN, obscuring medium with all values of column density $\nh$ simultaneously exist around AGN.
On the other hand, there are evidences that obscured and unobscured AGN present more similar $L_{\rm MIR}/L_{\rm X}$ ratios \citep[e.g.][]{Alonso-Herre01,Krabbe01,Lutz04,Horst06,Horst08} than that predicted by traditional torus models assuming a smooth distribution of dusty obscuring medium.
Recent works \citep{Nenkova08a,Nenkova08b,Hönig10a,Hönig10b} show that the distribution of dusty obscuring medium is clumpy rather than contiguous.
So, when we observe an AGN from one direction, both diffuse and reflected emission from all $\nh$ can be observed, in addition to the attenuated incident emission through an obscuring medium with a column density $\nh$ of this direction.
For this reason, we have made a modification to the original torus emission model of \cite{Ballantyne06b}.
The SED of an AGN with column density $\nh$ is constructed by adding the diffuse and reflected emission averaged over all 10 models with different $\nh$ to the attenuated incident emission through a particular $\nh$.
The weights are given by the probability distribution of the column densities, which is a function of $f_2({\rm log}L_{\rm X},z)$ (or covering factor) as discussed in Section \ref{ssect:f2}.
The SEDs constructed this way is called `unified SEDs'.

\begin{figure}
  \begin{center}
    \includegraphics[scale=0.69]{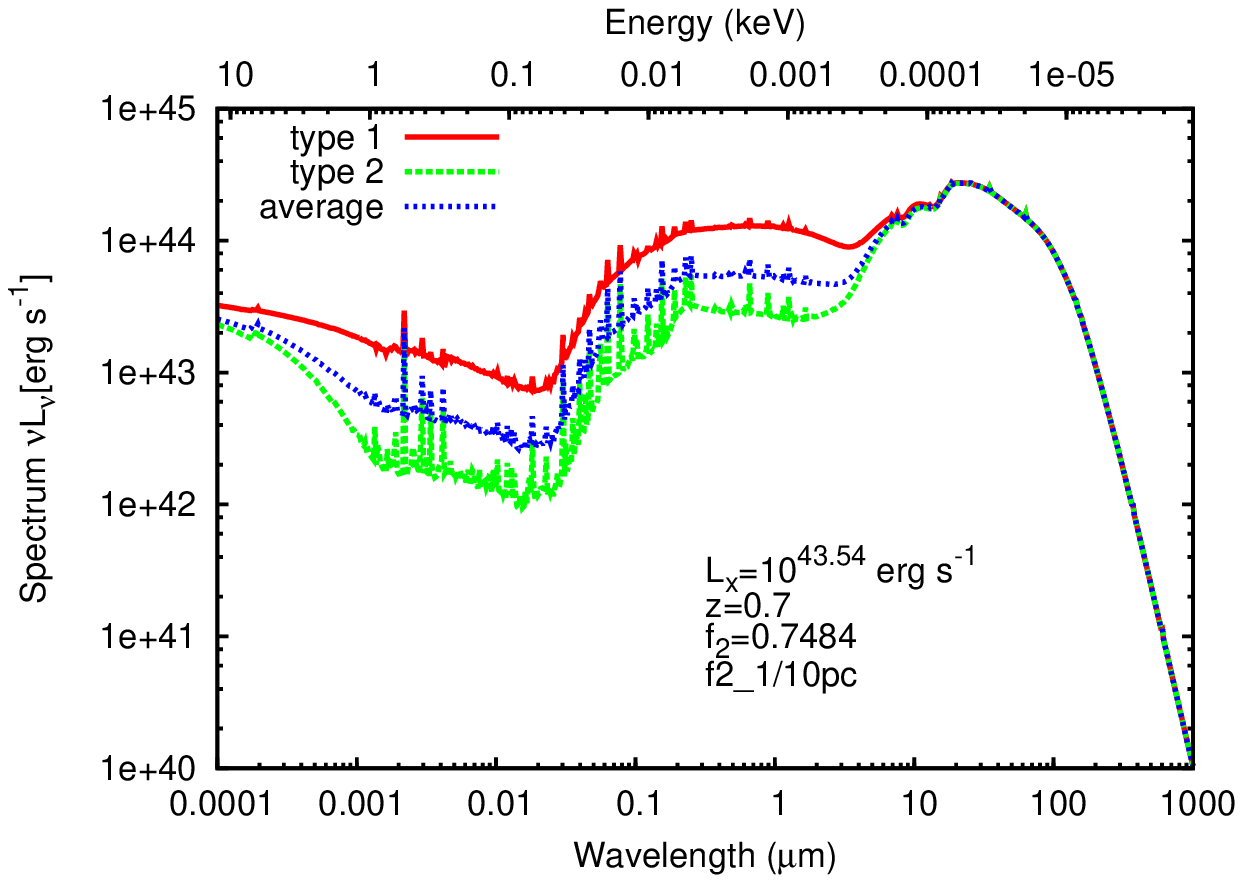}
    \includegraphics[scale=0.69]{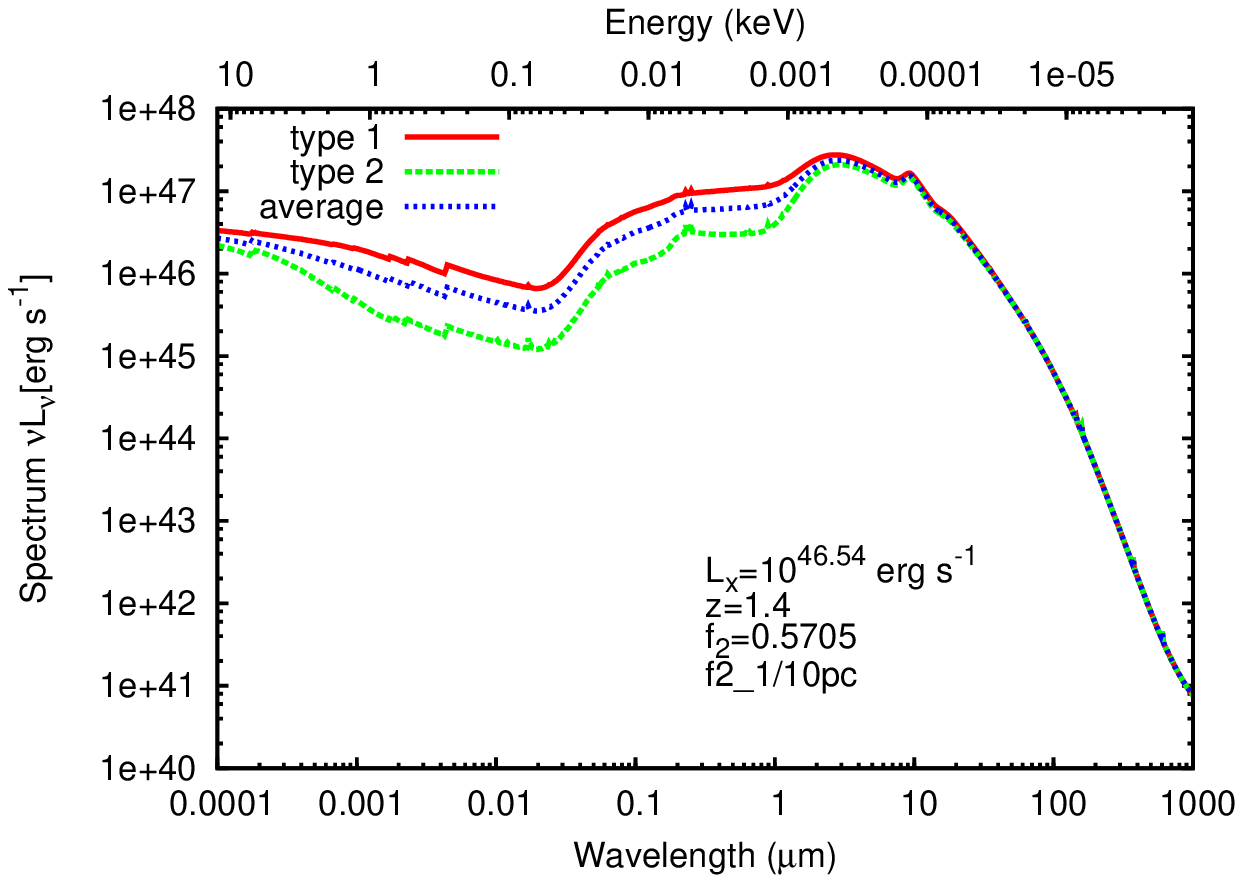}
  \end{center}
  \caption{Top: Rest-frame SEDs for a Seyfert-like AGN ($L_{\rm X}=10^{43.5}{\rm erg}~{\rm s}^{-1}$) at $z=0.7$ and obscured by dusty medium with an inner radius $10~\rm{pc}$, hydrogen density $n_{\rm{H}}=10^{4} \rm{cm}^{-3}$ and covering factor $f_2=0.7484$. The type-1 SED is shown in red, the type-2 SED is shown in green, while the average SED is shown in blue. Bottom: As top, but for a quasar-like AGN ($L_{\rm X}=10^{46.5}{\rm erg}~{\rm s}^{-1}$) at $z=1.4$ and obscured by dusty medium with a covering factor $f_2=0.5705$. These spectrums are taken from the `f2\_1' evolutionary grid (eq. \ref{eq:f2_1}).}
  \label{fig:SED}
\end{figure}

Finally, the `unified SEDs' undergo an average over $\nh$ again to produce the `$\nh$-averaged SEDs', which will be used to predict the IRLFs of AGN later.
Here, three types of $\nh$-averaged SEDs are constructed.
The type-1 SED is an average of the `unified SEDs' with $10^{20.0}\rm{cm}^{-2}\le\nh<10^{22}\rm{cm}^{-2}$.
The type-2 SED is an average of the `unified SEDs' with $10^{22}\rm{cm}^{-2}\le\nh\le10^{24.5}\rm{cm}^{-2}$.
The average SED is an average of the `unified SEDs' with $10^{20.0}\rm{cm}^{-2}\le\nh\le10^{24.5}\rm{cm}^{-2}$.
As an example, Fig \ref{fig:SED} shows the rest-frame SEDs taken from the `f2\_1' evolutionary grid (eq. \ref{eq:f2_1}) for a Seyfert-like AGN ($L_{\rm X}=10^{43.54}{\rm erg}~{\rm s}^{-1},z=0.7,f_2=0.7484$) and quasar-like AGN ($L_{\rm X}=10^{46.54}{\rm erg}~{\rm s}^{-1},z=1.4,f_2=0.5705$), respectively.

\subsection{Testing model SEDs of AGN}
\label{sssect:test}
The method we have used to model the SEDs of AGN is similar to that of \cite{Ballantyne06b}.
They have been extensively tested this method against large samples of AGNs.
Here, we present two additional tests that are more directly related to the goal of this paper, i.e. prediction of IRLFs from HXLF.
For this goal, the most important thing is correct X-ray to IR relation.

Recently, \cite{Gandhi09} found a strong mid-infrared:X-ray (${\rm log} \lambda L_{\lambda}({\rm 12.3 \mu m})$-${\rm log} L_{\rm 2-10 keV}$) luminosity correlation for a sample of local Seyferts, the cores of which have been resolved in the mid-IR.
The relation is given by
\begin{equation}
  \label{Lx_LIR1}
{\rm log} \lambda L_{\lambda}({\rm 12.3 \mu m})=-4.37+1.106~{\rm log} L_{\rm 2-10 keV},
\end{equation}
and is found to be valid in a wide range of luminosity and may extend into the quasar regime.
\cite{Mullaney11} converted this correlation to that between ${\rm log} L_{\rm IR}$ and ${\rm log} L_{\rm 2-10 keV}$, which is given by
\begin{equation}
  \label{Lx_LIR2}
  {\rm log} \frac{L_{\rm IR}}{10^{43} {\rm erg~s^{-1}}} = 0.53 + 1.11~{\rm log} \frac{L_{\rm 2-10 keV}}{10^{43}{\rm erg~s^{-1}}}.
\end{equation}

\begin{figure}
  \begin{center}
    \includegraphics[scale=0.8]{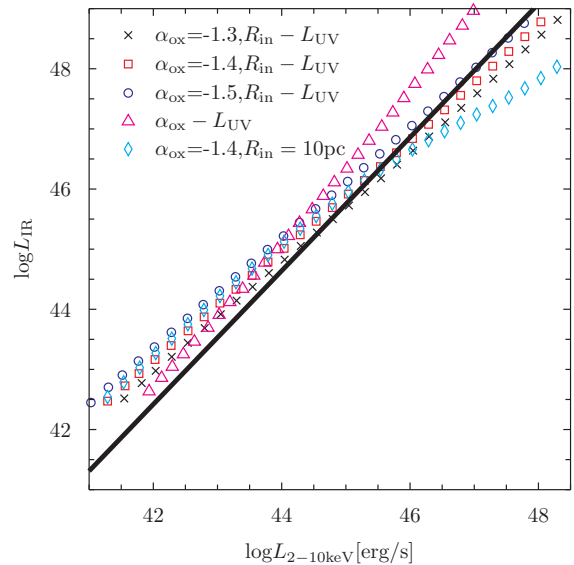}
  \end{center}
  \caption{Test of the  ${\rm log} L_{\rm IR}$-${\rm log} L_{\rm 2-10 keV}$  relation computed from our model SEDs of AGN against the observational relation (black-solid line) given  by \citet{Mullaney11}. The results obtained by using different choices of $\alpha_{\mathrm{ox}}$ and $R_{\mathrm{in}}$ are presented, respectively. The $L_{\rm UV}$-dependent $R_{\rm in}$ is given by Eq. \ref{eq:Rin}, while the $L_{\rm UV}$-dependent $\alpha_{\mathrm{ox}}$ is given by Eq. \ref{eq:alphaox}.}
  \label{fig:Lx_vLv}
\end{figure}

In Fig \ref{fig:Lx_vLv}, the  ${\rm log} L_{\rm IR}$-${\rm log} L_{\rm 2-10 keV}$ relation computed from our model SEDs of AGN by using different choices of $\alpha_{\mathrm{ox}}$ and $R_{\mathrm{in}}$ are tested against the observational relation given by \cite{Mullaney11}.
As shown in the figure, the result obtained for $\alpha_{\mathrm{ox}}=-1.4$ and $R_{\mathrm{in}}=10~{\rm pc}$ (as in \citet{Ballantyne06b}) significantly deviates from the nearly linear relation given by \citet{Mullaney11}.
This problem has also been noticed by \cite{Draper11}.
We found that a more linear ${\rm log} L_{\rm IR}$-${\rm log} L_{\rm 2-10 keV}$ relation can be obtained if $R_{\mathrm{in}}$ decreases with $L_{\rm UV}$ (as given by Eq. \ref{eq:Rin}).
With the typical value of $\alpha_{\mathrm{ox}}=-1.4$, this lead to a result similar to that of \cite{Mullaney11}, especially at the high-luminosity range.
At the low-luminosity range, a larger $\alpha_{\mathrm{ox}}$ seems required.
However, when the $\alpha_{\mathrm{ox}}$-$L_{\rm UV}$ relation given by \cite{Hopkins07a} (as given by Eq. \ref{eq:alphaox}) is assumed, a too steep relation is obtained.
So, these results support the anticorrelation between $\alpha_{\mathrm{ox}}$ and $L_{\rm UV}$ found by other independent observations \citep[e.g.][]{Steffen06,Vagnetti10}, but imply a more flat relation.

\section{CONNECTING X-RAY AND IR}
\label{sect:connect}
The IR is less affected by the selection effects due to obscuration suffered by optical and X-ray, while X-ray is currently the most efficient for selecting AGN to high redshifts.
Connecting X-ray and IR can provide a more clear view on the evolution of AGN populations.
If the evolution of AGN shown in HXLF and $f_2$, which are revealed mainly from X-ray observations, are intrinsic, they should be shown somehow in the directly observed IRLFs of AGN.

\subsection{The mid-IR LFs of type-1 and/or type-2 AGN}
\label{ssect:irlf_z}
Since the HXLF tells us how the number density of AGN per increment of ${\rm log} L_{\rm X}$ changes with $z$ and $L_{\rm X}$, the following expression can be used to relate HXLF to IRLFs ${\rm d}\Phi/{\rm d}({\rm log} \nu L_{\nu})$:
\begin{equation}
\label{eq:irlf}
\frac{{{\rm d}{\Phi}}}
{{{\rm d}({\rm log} \nu {L_\nu })}} = \frac{{{\rm d}\Phi }}
{{{\rm d}({\rm log} {L_{\rm X}})}}\frac{{{\rm d}({\rm log} {L_{\rm X}})}}
{{{\rm d}({\rm log} {\nu }{L_\nu })}}.
\end{equation}
Here, ${{{\rm d}\Phi }}/{{{\rm d}({\rm log} {L_{\rm X}})}}$ is the HXLF of AGN given in Section \ref{ssect:hxlf}, and ${L_\nu }$ is the luminosity at a given wavelength.

In Section \ref{sect:seds}, we have obtained the SEDs spanning from X-ray all the way to IR for AGN with different luminosities and redshifts.
So, the dependence of IR luminosity on hard X-ray luminosity, as described by ${{{\rm d}({\rm log} {L_{\rm X}})}}/{{{\rm d}({\rm log} {\nu }{L_\nu })}}$, can be obtained from the SEDs easily.
For predicting the IRLFs for total AGN (type-1 + type-2), we use the average SED presented in Section \ref{sect:seds}.
The IRLF of AGN is not an integrated quantity, and so much more sensitive to the evolution trends in the HXLF and the $f_2$ than the cumulative number count distribution and background spectra intensity of AGN.
Using the IRLFs for total AGN has the advantage of being independent of the methods used to do a further classification of AGN, such as detailed optical emission line spectra, or an accurate measurement of X-ray absorbing column densities $\nh$.
For these reasons, \cite{Ballantyne06b} suggested to use IRLFs for total AGN to distinguish different evolution models of AGN obscuration.
However, the results of \cite{Ballantyne06b} showed that the IRLFs is not very sensitive to the evolution model of AGN obscuration unless at much longer IR wavelengths where the contaminant from star formation is important.

The separated IRLFs for type-1 and type-2 AGN respectively are expected to be much more sensitive to the overall evolution of AGN spatial density and obscuration, although a detailed classification are required.
The classification of AGN into type-1 and type-2 involves with the problem of consistency between different classification methods.
However, this may be a possible key to the problem of the evolution of AGN obscuration, since inconsistent classification methods will directly result in very different conclusions for the evolution of AGN obscuration.
By separating the IRLFs to that for type-1 and type-2 AGN, the intrinsic evolution of AGN and the variations just resulted from the evolution of AGN obscuration, can be investigated in more detail and possibly clarified.
Furthermore, by considering the IRLFs for type-1 and type-2 AGN separately, the modeling of AGN SEDs can be further constrained.
So, it would be more fruitful to separate the IRLFs of AGN into that for type-1 and type-2, respectively.
This may provide a more useful tool to explore the properties of obscuring medium around different types of AGN.

The separated IRLFs of type-1 and type-2 AGN are given as,
\begin{equation}
\label{eq:irlf1}
\frac{{{\rm d}{\Phi _1}}}
{{{\rm d}({\rm log} \nu {L_\nu })}} = \frac{{{\rm d}\Phi }}
{{{\rm d}({\rm log} {L_{\rm X}})}}(1 - {f_2}({\rm log} {L_{\rm X}},z))\frac{{{\rm d}({\rm log} {L_{\rm X}})}}
{{{\rm d}({\rm log} {\nu }{L_\nu })}},
\end{equation}
and
\begin{equation}
\label{eq:irlf2}
\frac{{{\rm d}{\Phi _2}}}
{{{\rm d}({\rm log} \nu {L_\nu })}} = \frac{{{\rm d}\Phi }}
{{{\rm d}({\rm log} {L_{\rm X}})}}{f_2}({\rm log} {L_{\rm X}},z)\frac{{{\rm d}({\rm log} {L_{\rm X}})}}
{{{\rm d}({\rm log} {\nu }{L_\nu })}},
\end{equation}
where ${f_2}({\rm log} {L_{\rm X}},z)$ is the fraction of type-2 AGN.
We use the type-1 SED presented in Section \ref{sect:seds} to predict the IRLFs of type-1 AGN, while using the type-2 SED to predict the IRLFs of type-2 AGN.

\section{RESULTS}
\label{sect:irlf_result}
In this section, we present the predicted mid-IR LFs for total, type-1 and type-2 AGN, respectively.
The SEDs of AGN used to obtain the X-ray to IR luminosity relation are computed by assuming a constant $\alpha_{\mathrm{ox}}=-1.4$, and $L_{\rm UV}$-dependent $R_{\rm in}$ as described by Eq. \ref{eq:Rin}.
We leave the discussion of $L_{\rm UV}$-dependent $\alpha_{\mathrm{ox}}$ in Section \ref{sect:discu}.
The results for all combinations of HXLF as given in Section \ref{ssect:hxlf} and different evolution models of AGN obscuration as given in Section \ref{ssect:f2} are presented and then compared with the measurements of mid-IR LFs of AGN currently available, respectively.
Since we are mainly interested in finding out much obvious trends, a simple qualitative comparison by eye rather than a much detailed fitting \footnote{Much more careful considerations of the covariance between points or systemic errors are not included as well.}, is taken here.

\subsection{The mid-IR LFs of total AGN}
\label{ssect:irlfsum}

\begin{figure*}
  \begin{center}
    \includegraphics[scale=0.6]{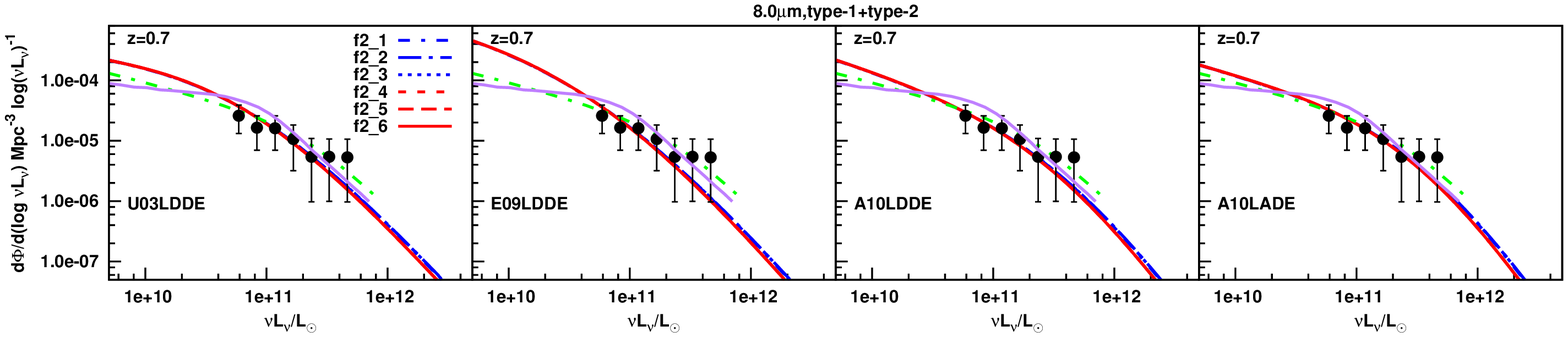}
  \end{center}
  \caption{Rest-frame $8.0$ \micron~LF for total AGN at $z=0.7$ as predicted from the LDDE modeling of HXLF of \citet{Ueda03}, \citet{Ebrero09}, \citet{Aird10}, and the LADE modeling of HXLF of \citet{Aird10}, respectively. At each panel, the results for six evolution models of  $f_2$ are presented. The blue lines show the results for the three evolution models of $f_2$ given by \citet{Ballantyne06a}, who constrained the evolution of $f_2$ by fitting to the shape of CXRB spectrum and X-ray number counts. The short dashed-doted, long dashed-doted, and doted lines show the results for the `f2\_1', `f2\_2', and `f2\_3' evolution models, respectively. The red lines show the results for the three evolution models of $f_2$ given by \citet{Hasinger08} from direct X-ray observations of the evolution of type-2 AGN fraction. Here, the short-dashed, long-dashed, and solid lines show the results for the `f2\_4', `f2\_5', and `f2\_6' evolution models, respectively. The data points are the $8.0$ \micron~IRS-decomposed AGN LF of \citet{Fu10} at $z\sim0.7$. The green dot-dashed lines show the obscuration-corrected AGN bolometric LF of \citet{Hopkins07a} (taken from \citet{Fu10}). The purple solid lines show the LF for total AGN of \citet{Matute06} as combined and converted to $8.0$ \micron~at $z\sim0.7$ by \citet{Fu10}.}
  \label{fig:irlfsum_8}
\end{figure*}
\begin{figure*}
  \begin{center}
    \includegraphics[scale=0.6]{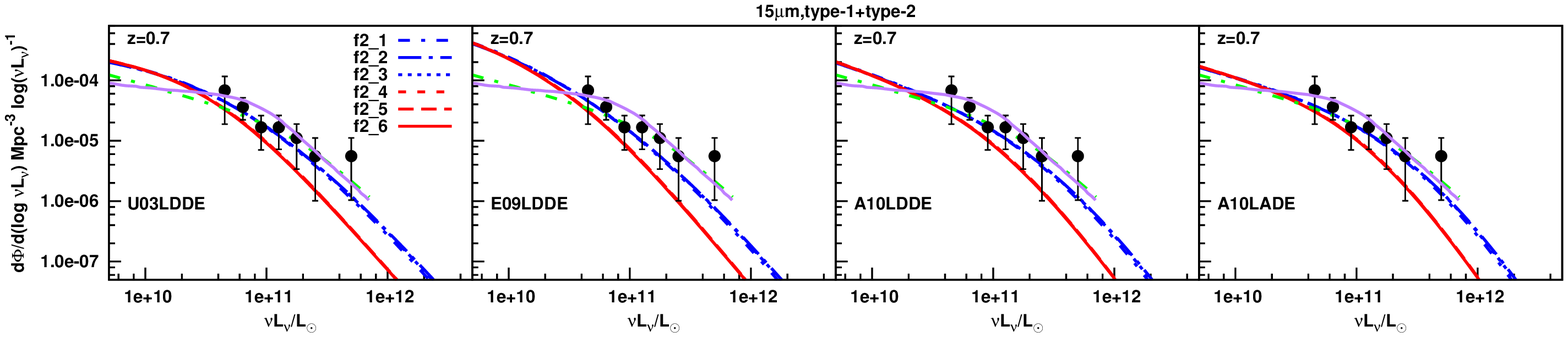}
  \end{center}
  \caption{Similar to Fig. \ref{fig:irlfsum_8}, but for the $15$ \micron~LF. The AGN obscuration evolution models of \citet{Ballantyne06b} give better agreements with the measurements of \citet{Fu10}, \citet{Matute06}, and the results of \citet{Hopkins07a}.}
  \label{fig:irlfsum_15}
\end{figure*}

In Figs. \ref{fig:irlfsum_8} and \ref{fig:irlfsum_15}, the predicted rest-frame $8.0$ and $15$~\micron~LF for total AGN are shown.
In each panel, the results are predicted from an evolution model of HXLF and six evolution models of AGN obscuration as discussed in Section \ref{ssect:f2}.
The observational results used for comparison are from \cite{Fu10}.
They used high-quality \spitzer~$7-38$~\micron~spectra to cleanly separate star formation and AGN in individual galaxies for a $24$~\micron~ flux-limited sample of galaxies at $z\sim0.7$, and decomposed the mid-IR LFs between star formation and AGN.

As can be clearly seen in Figs. \ref{fig:irlfsum_8} and \ref{fig:irlfsum_15}, our results agree with that of \cite{Fu10}, \cite{Matute06} and \cite{Hopkins07a} reasonably.
These general agreements show that the methods we have used to model the SEDs of AGN and to predict corresponding mid-IR LFs from HXLF are basically reasonable.
Specifically, different evolution models of HXLF give very similar results at $8.0~\rm{\mu m}$ and $15~\rm{\mu m}$.
However, different evolution models of AGN obscuration are distinguishable at $15~\rm{\mu m}$, while not at $8.0~\rm{\mu m}$.
As shown in Fig \ref{fig:irlfsum_15}, the results at $15~\rm{\mu m}$ are divided into two groups, corresponding to using models from \cite{Ballantyne06a} and using models constructed according to recent results of \cite{Hasinger08}, respectively.
The results predicted by using the evolution models of AGN obscuration from \cite{Ballantyne06a} are in a better agreement with the measurements of \cite{Fu10} and the results from other authors, especially at the relative higher luminosities.
It seems that the evolution of AGN obscuration are better described by the models of \cite{Ballantyne06a} at the redshift and luminosity ranges covered by the measurements of \cite{Fu10}, i.e. $z\lesssim1$ and $\nu L_{\nu}(8.0~\rm{\mu m},15~\rm{\mu m})<10^{12}\Lsun$.

\subsection{The mid-IR LFs of type-1 AGN and type-2 AGN}
\label{ssect:irlf1_2}
As mentioned above, it is more fruitful to separate the mid-IR LFs to that for type-1 and type-2 AGN, respectively.
Here, we present the mid-IR LFs for type-1 and type-2 AGN and then compare them with the mid-IR observational results of \cite{Brown06} and \cite{Matute06}, respectively.

\subsubsection{The $8.0$ \micron~LF of type-1 AGN}
\label{sssect:irlf1_8}
From a sample consisting of 292 $24$ \micron~sources brighter than 1 mJy selected from \spitzer~MIPS survey, \citet{Brown06} have determined the rest-frame $8.0$ \micron~LF for type-1 quasars with $1<z<5$ and $1.5<z<2.5$, respectively.
\cite{Ballantyne06b} used these results (in their Fig. 13), but compared them with the predicted mid-IR LFs for total AGN.
Despite this, they found that the predicted and measured LFs show a surprising agreement.
As suggested by \cite{Brown06}, if the fraction of obscured quasars decreases rapidly with increasing luminosity, the type-1 quasar LF given by them would appropriate the LF of all quasars at the highest luminosities.
However, if there are indeed very few type-2 AGN at very high luminosities, and the type-1 AGN LF provide good approximation of the total LF at the high luminosities, then this would be an important constraint for the evolution of AGN obscuration at high luminosities.
This important information was not fully utilized in the method of \cite{Ballantyne06b}.
So, it would be more reasonable and fruitful to predict the mid-IR LFs for only type-1 AGN from different evolution models of HXLF and obscuration of AGN, and then compare them with the measurements of \cite{Brown06}.

\begin{figure*}
  \begin{center}
    \includegraphics[scale=0.6]{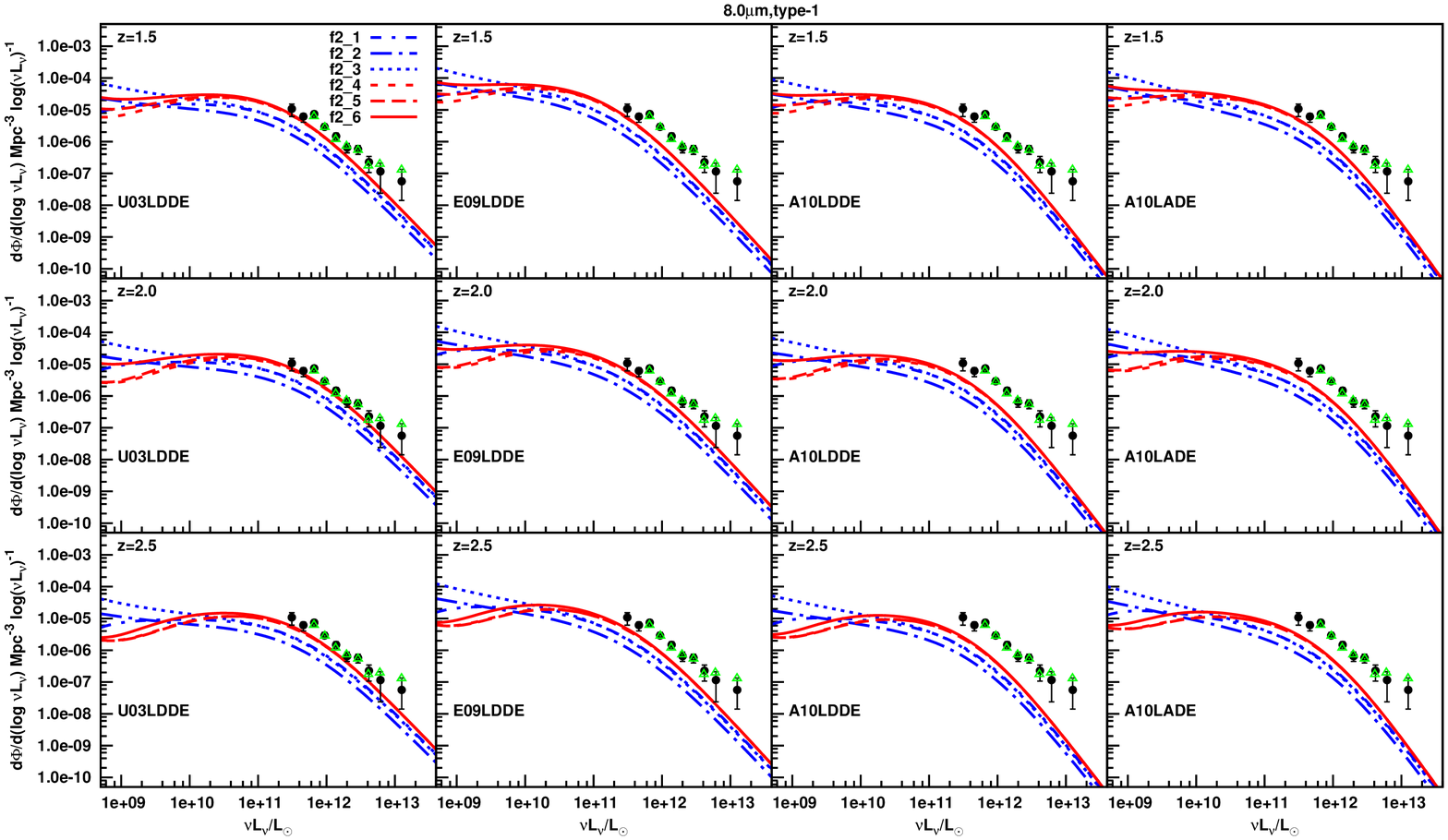}
  \end{center}
  \caption{Rest-frame $8.0$ \micron~LF for type-1 AGN at $z=1.5, 2.0$, and $2.5$ as predicted from the LDDE modeling of HXLF of \citet{Ueda03}, \citet{Ebrero09}, \citet{Aird10}, and the LADE modeling of HXLF of \citet{Aird10}, respectively. At each panel, the results for six evolution models of  $f_2$ are presented with line styles and colours that are the same as in Fig. \ref{fig:irlfsum_8}. The data points show the measured $8.0$ \micron~LF of type-1 quasars as determined by \citet{Brown06} from a sample consists of 292 $24$ \micron~sources brighter than 1 mJy and selected from \spitzer~MIPS survey. The black solid points denote the result for objects over the redshift range $1<z<5$, and the green triangles for those with $1.5<z<2.5$. All the results predicted from HXLF, especially those recently presented, tend to underestimate the number of IR-luminous AGN, independent of the choices of the evolution of HXLF and obscuration.}
\label{fig:irlf1_8}
\end{figure*}
In Fig. \ref{fig:irlf1_8}, we present the predicted rest-frame $8.0$ \micron~LF of type-1 AGN at $z=1.5, 2.0$, and $2.5$, and compare them with the measurements of \cite{Brown06}.
As expected, the results predicted from different evolution models of AGN obscuration are more distinguishable when the rest-frame $8.0$ \micron~LF for only type-1 AGN, instead of total AGN, are used.
The rest-frame $8.0$ \micron~LF of type-1 AGN predicted from different choices of HXLF tend to underestimate the number of AGN as measured by \cite{Brown06}, and no matter which evolution model of AGN obscuration is used.
Surprisingly, using the HXLF of \cite{Ueda03} result in a better agreement with the measurements of \cite{Brown06} than  using all the more recent HXLF measurements.
The results shown here may have, to some extent, confirmed the credibility of widely used results of \cite{Ueda03}.
However, this may just be a coincidence, since more recent observational determination of HXLF are generally expected to be more accurate.
More reasonable conclusions can only be obtained from comparisons with other independent measurements.

\subsubsection{The $15$ \micron~LF of type-1/type-2 AGN}
\label{sssect:irlf_15}
\begin{figure*}
  \begin{center}
    \includegraphics[scale=0.6]{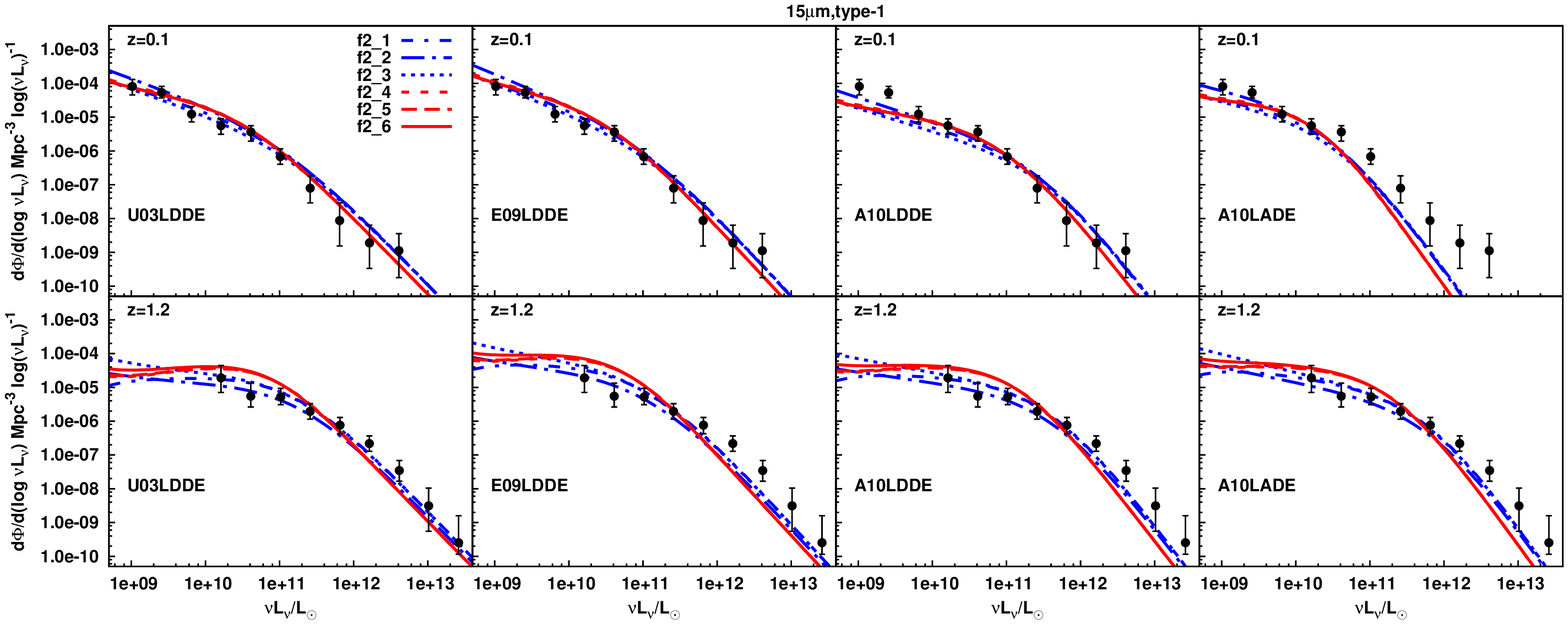}
  \end{center}
  \caption{Rest-frame $15$ \micron~LF for type-1 AGN at $z=0.1$, and $1.2$ as predicted from the LDDE modeling of HXLF of \citet{Ueda03}, \citet{Ebrero09}, \citet{Aird10}, and the LADE modeling of HXLF of \citet{Aird10}, respectively. The data points are the measured $15$ \micron~LF of type-1 AGN determined by \citet{Matute06} from a sample of type-1 AGN with redshift in $z=[0,0.2]$ (top) and $z=[0.2,2.2]$ (bottom) selected at $15$ \micron~(ISO) and $12$ \micron~(IRAS), and classified by their optical spectra. At each panel, the results for six evolution models of $f_2$ are presented with line styles and colours as in Fig. \ref{fig:irlfsum_8}. The results predicted from all HXLFs tend to underestimate the number of the most IR-luminous AGN at $z=1.2$. However, this is not the case at $z=0.1$, unless the LADE modeling of HXLF of \citet{Aird10} is used.}
  \label{fig:irlf1_15}
\end{figure*}
\begin{figure*}
  \begin{center}
    \includegraphics[scale=0.6]{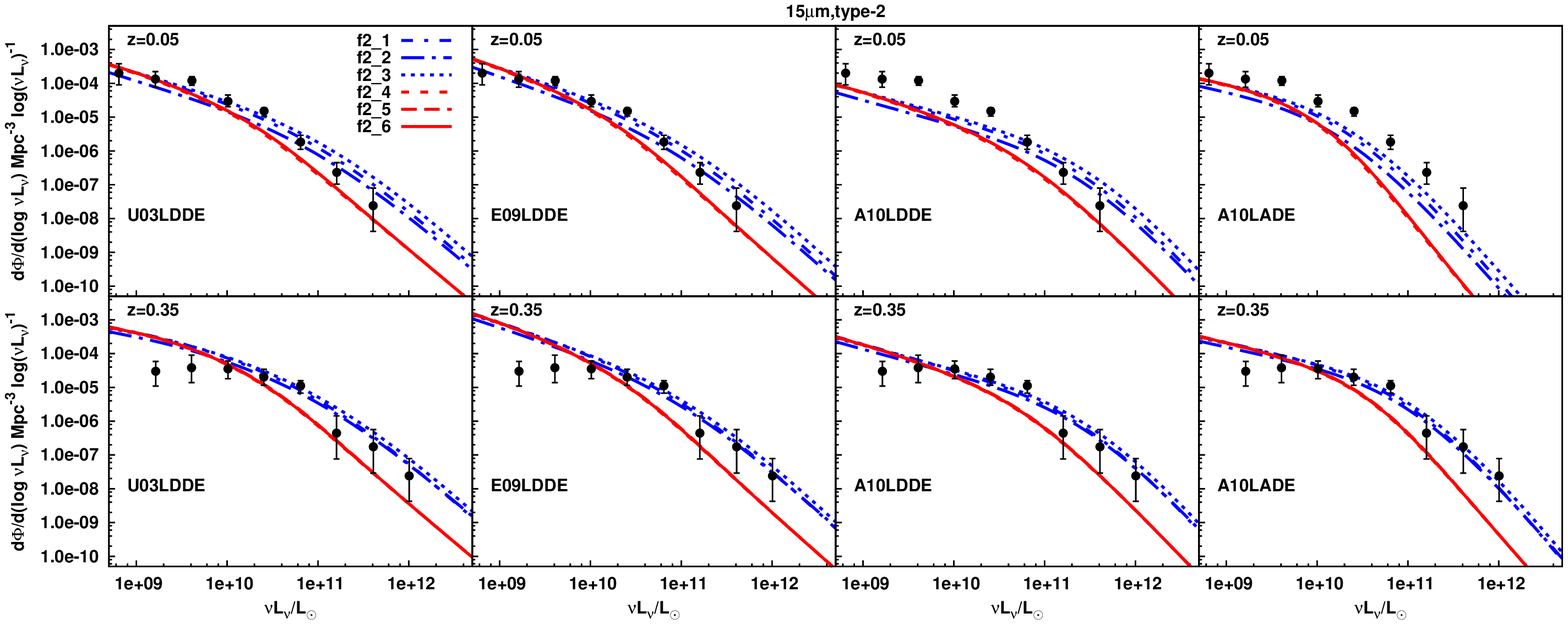}
  \end{center}
  \caption{Similar to Fig. \ref{fig:irlf1_15}, but for type-2 AGN at $z=0.05$, and $0.35$. The data points are the measured $15$ \micron~LF of type-2 AGN determined by \citet{Matute06} from a sample of type-2 AGN with redshift in $z=[0,0.1]$ (top) and $z=[0.1,0.6]$ (bottom) selected at $15$ \micron~(ISO) and $12$ \micron~(IRAS), and classified by their optical spectra. The measurement of the mid-IR LF of type-2 AGN is much poorer than that of type-1 AGN, and therefore, is much harder to be explained. However, the results predicted from the AGN obscuration evolution model constructed according to the results of \citet{Hasinger08} are below even the current measurements.}
  \label{fig:irlf2_15}
\end{figure*}

From a sample of AGN selected at $15$ \micron~(ISO) and $12$ \micron~(IRAS), \cite{Matute06} measured the rest-frame $15$ \micron~LF of type-1 and type-2 AGN, which are classified based on their optical spectra, separately. 
In Figs. \ref{fig:irlf1_15} and \ref{fig:irlf2_15}, we show the predicted rest-frame $15$ \micron~LF for type-1 AGN and type-2 AGN and compare them with the results of \cite{Matute06}.

Fig. \ref{fig:irlf1_15} presents the rest-frame $15$ \micron~LF of type-1 AGN at $z=0.1$ and $1.2$.
As can be clearly seen, the predicted results show reasonable agreements with the measurements of \cite{Matute06}.
However, it is also clear,  especially at $z=1.2$, that the predicted IRLFs tend to underestimate the number of the most IR-luminous type-1 AGN, which is independent of the choices of the evolution of HXLF and obscuration.
Interestingly, the measurements at $z=0.1$ can be basically explained by the results predicted from most HXLFs.
The only exception is the result predicted from the HXLF of \cite{Aird10} modeled with LADE, which significantly underestimated the number of the most IR-Luminous AGN even at $z=0.1$.
Similar to that shown in Fig. \ref{fig:irlf1_8}, these results show that the mid-IR LFs predicted from HXLF tend to  underestimated the number of the most IR-luminous AGN, and become significant at $z\gtrsim1$.

Fig. \ref{fig:irlf2_15} presents the rest-frame $15$ \micron~LF of type-2 AGN at $z=0.05$ and $0.35$.
As mentioned by \cite{Matute06}, the observational determination of the rest-frame $15$ \micron~LF of type-2 AGN is much poorer than that of type-1 AGN.
So, their measurement of the density of type-2 AGN can only be considered as a lower limit.
However, as can be seen in Fig. \ref{fig:irlf2_15}, the results predicted from the AGN obscuration evolution models that are constructed according to the results of \citet{Hasinger08} tend to underestimate even the number of type-2 AGN currently measured.
Due to the much bigger uncertainties in the measurements of the 15 \micron~LF of type-2 AGN, more definitive conclusions cannot be drawn.

\section{DISCUSSION}
\label{sect:discu}
By separating the mid-IR LFs of AGN to that for type-1 and type-2 AGN respectively, the modeling of AGN SEDs, the evolution of LFs and obscuration of AGN can be further constrained.
The results presented in the Section \ref{sect:irlf_result} show that the mid-IR LFs predicted from HXLF tend to underestimate the number of the most IR-luminous AGN, despite of the general agreements between predictions and measurements.
This is independent of the choices of the evolution models of HXLF and obscuration of AGN, and even more obvious for the HXLFs recently proposed.
Meanwhile, this trend seems not significant for AGN with $z\lesssim1$ and/or less luminous at IR.

Here, we discuss some possible explanations for this contradiction between HXLFs and mid-IR LFs.
Firstly, this may be caused by the missing fraction of AGN, especially those heavily obscured Compton-thick AGN that cannot be detected by current X-ray observations.  
Recently, \cite{Fu10} compared their mid-infrared spectroscopic selection with other AGN identification methods and concluded that only half of the mid-infrared spectroscopically selected AGN were detected in X-ray.
However, after considering this we find that it only result in a slight improvement to the prediction of IRLFs from HXLF.
Furthermore, this explanation needs a larger fraction of missing AGN at the high-luminosity end, which is in contradiction with the general expectation that AGN dominate in the most IR-luminous sources.

Secondly, the contribution of star formation in AGN host to mid-IR, which has not been considered yet, may be important.
If this is important, the X-ray to mid-IR relation used to predict mid-IR LFs from HXLFs in Section \ref{sect:irlf_result} needs to be corrected significantly. 
This is particularly important for sources that are not spatially resolved, or with intensive star formation near the nuclear region \citep{Lutz04,Horst08}.
We find that if the contribution of star formation in the host to $8.0~\rm{\mu m}$ and $15~\rm{\mu m}$ emission are comparable to the reprocessed nuclear emission, the $8.0~\rm{\mu m}$ and $15~\rm{\mu m}$ LF predicted from HXLF can be consistent with corresponding mid-IR measurements.
Currently, it is still difficult to separate the contribution of star formation and AGN to the IR emission of galaxies, especially in systems where the two are comparable and their additive effects are non-linear \citep{Hopkins10b}.
Especially, the relative fractions of their contributions to mid-IR are likely to be different in different kinds of galaxies, and may change with both luminosity and redshift of the source.
However, even for powerfully star-forming quasars, the contribution of star formation to mid-IR is small \citep{Netzer07-2}.
So, the possible contribution of star formation to mid-IR is not likely the main reason for the contradiction.

Thirdly, the contradiction found in Section \ref{sect:irlf_result} may represent limitations in the torus model used so far.
Although the simple torus model inherited from \cite{Ballantyne06b} has been well tested, the distribution and composition of the obscuring medium around AGN are still very uncertain.
Meanwhile, this CLOUDY based torus model essentially assume a smooth distribution of dusty obscuring medium.
Recently, a clumpy distribution of dusty obscuring medium is suggested by some authors \citep{Nenkova02,Hönig06}.
These authors have recently proposed sophisticated clumpy torus models \citep{Nenkova08a,Nenkova08b,Hönig10a,Hönig10b} that are in a better agreement with current IR observations of AGN.
Unfortunately, these clumpy torus models mainly give the IR emission properties of AGN, while the self-consistent hard X-ray property is not presented.
To give the X-ray to mid-IR luminosity ratios that are more comparable to the observational results of \cite{Mullaney11}, we have made some improvements to the original torus model of \cite{Ballantyne06b}.
However, the improved torus model still have some limitations, which is worth additional efforts but is beyond the scope of this paper.

Finally, as shown in Section \ref{sssect:test}, the anticorrelation between $\alpha_{\mathrm{ox}}$ and $L_{\rm UV}$, which has been found by many observations, is important for giving  X-ray to mid-IR luminosity ratios that are more comparable to the result of \cite{Mullaney11}.
Here, we present the results obtained by assuming $\alpha_{\mathrm{ox}}=-1.5$ instead of the typical value of $-1.4$ as used in Section \ref{sect:irlf_result}.
\begin{figure*}
  \begin{center}
    \includegraphics[scale=0.6]{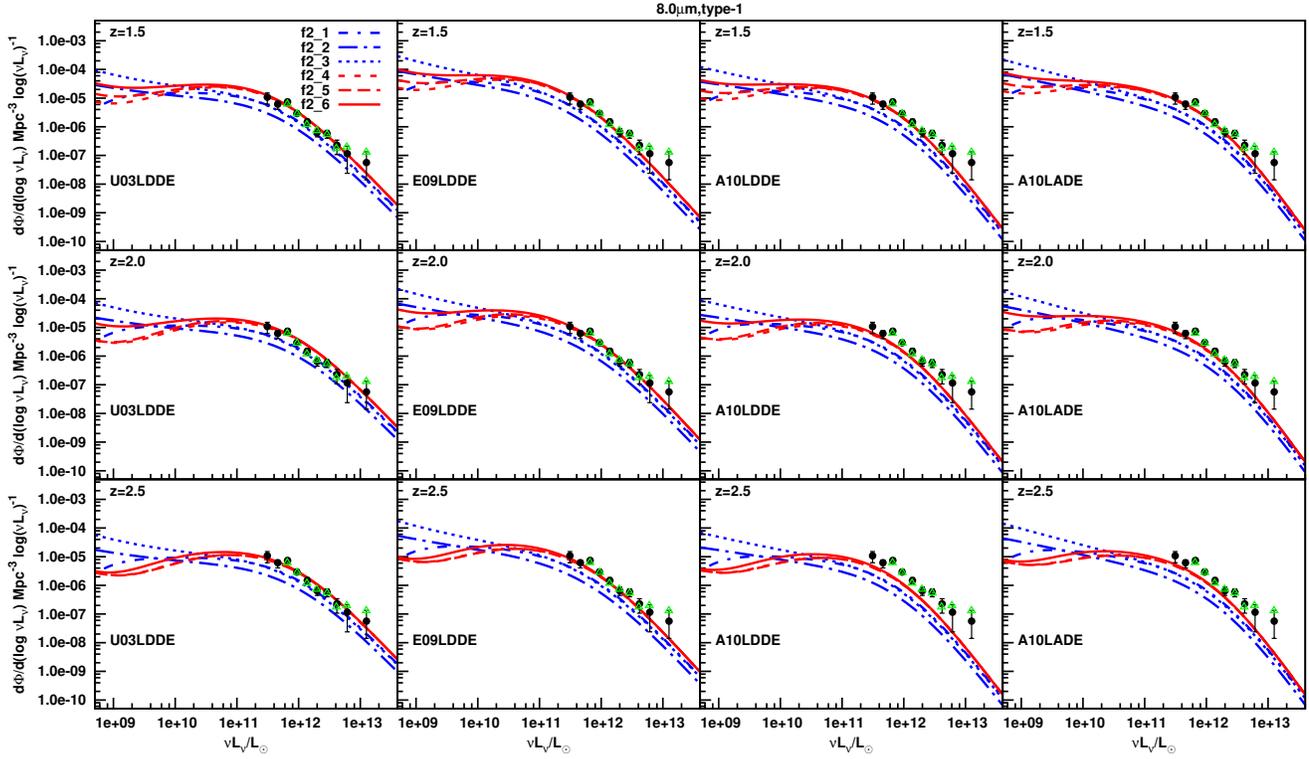}
  \end{center}
  \caption{Similar to Fig. \ref{fig:irlf1_8}, but now $\alpha_{\mathrm{ox}}$ is assumed to have a smaller value of $-1.5$ instead of the typical value of $-1.4$ as used in Section \ref{sect:irlf_result}. As can be seen, the results have been largely improved after this small change. Now, the measurements of \citet{Brown06} can be well explained by recently proposed HXLF, especially that of \citet{Ebrero09}, when combined with the obscuration evolution models constructed according to the results of \citet{Hasinger08}. An even smaller $\alpha_{\mathrm{ox}}$ seems required at $\nu L_{\nu}(8.0~\rm{\mu m})\gtrsim5*10^{12}\Lsun$.}
\label{fig:irlf1_8_corrected}
\end{figure*}
\begin{figure*}
  \begin{center}
    \includegraphics[scale=0.6]{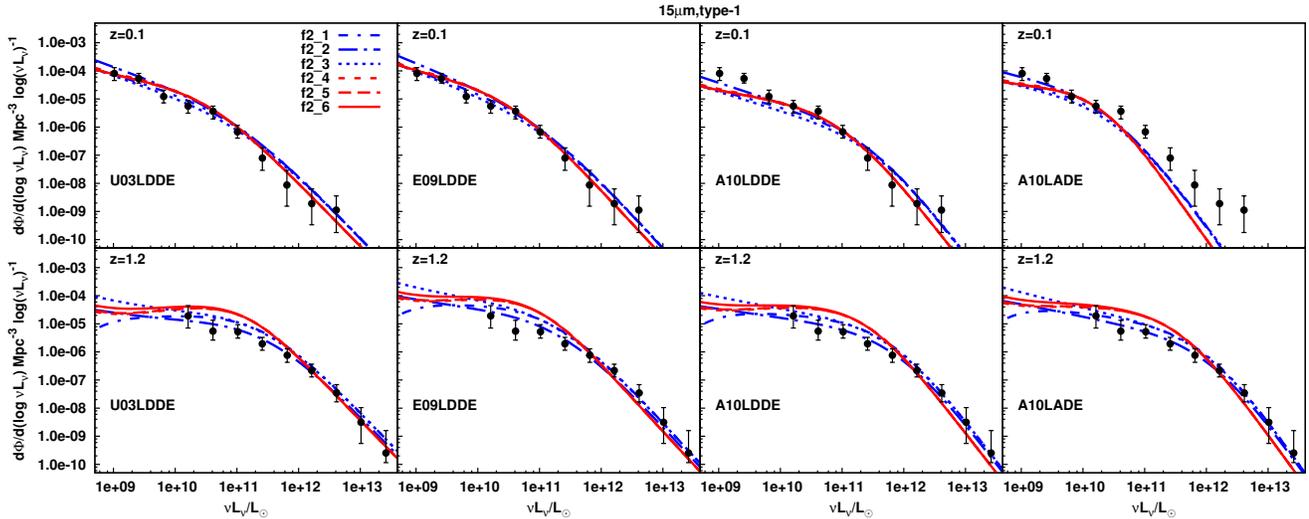}
  \end{center}
  \caption{Similar to Fig. \ref{fig:irlf1_15}, but now $\alpha_{\mathrm{ox}}$ is assumed to have a smaller value of $-1.5$ at $z=1.2$ instead of the typical value of $-1.4$ as used in Section \ref{sect:irlf_result}.  At $z=1.2$, it is clear that the results at highest luminosities have been largely improved. Now, the AGN obscuration evolution models proposed by \citet{Ballantyne06a} seems more favored. However, the change of $\alpha_{\mathrm{ox}}$ is not required at $z=0.1$, unless the LADE modeling of HXLF of \citet{Aird10} is used.}
\label{fig:irlf1_15_corrected_nh2}
\end{figure*}
As can be seen in Figs. \ref{fig:irlf1_8_corrected} and \ref{fig:irlf1_15_corrected_nh2},  the mid-IR LFs measurements of \cite{Brown06} and \cite{Matute06} can now be explained much better by the results predicted from the HXLFs recently proposed, especially that of \cite{Ebrero09}.
On the other hand, it is interesting to notice the dramatical difference between the results in Figs. \ref{fig:irlf1_8_corrected} and \ref{fig:irlf1_15_corrected_nh2}.
The $8.0~\rm{\mu m}$ LF measurement of \cite{Brown06} is for luminous quasars with $z>1$, while the $15~\rm{\mu m}$ LF of \cite{Matute06} is mainly for Seyferts at much lower luminosities and redshifts.
While the results at $8.0~\rm{\mu m}$ favor the obscuration evolution models constructed according to the results of \cite{Hasinger08}, the results at $15~\rm{\mu m}$ nevertheless give more supports to the models proposed by \cite{Ballantyne06a}.

These results imply that the obscuration of quasars are different from that of Seyferts.
Luminous quasars often associate with galaxy major mergers \citep{Canalizo01} or interactions \citep{Hutchings87,Disney95,Bahcall97,Kirhakos99}, while there are little observational evidences for less luminous Seyfert galaxies being associated with mergers \citep{Laurikainen95,Schmitt01,Grogin05}.
If the evolution and fueling mechanisms of quasars are very different from that of lower luminosity Seyfert galaxies, it is natural to expect that the distribution and evolution of the obscuring medium around them are very different.
As pointed out by \cite{Ballantyne06b}, the dusty mediums obscuring luminous quasars are likely distributed in a larger scale and linked to the starburst region, while lower luminous quasars and Seyferts are obscured by commonly suggested compact torus located at much smaller scale.

\section{SUMMARY}
\label{sect:summary}
We have presented a detailed comparison between the $2-10$ keV HXLFs and mid-IR LFs of AGN.
The combination of hard X-ray and mid-IR provide complementary views for understanding the evolution of LFs and obscuration of AGN and their co-evolution with galaxies.
Four measurements of the HXLF of AGN have been collected from the literatures for comparison.
A simple but well tested torus model, which is based on photoionization and radiative transfer code CLOUDY, is then employed to model the composite X-ray to IR SEDs for AGN with different luminosities and redshifts.
In the modeling of SEDs, we have assumed six evolution models of AGN obscuration, which are constrained by the CXRB \citep{Ballantyne06a}, or constructed according to recent direct measurement \citep{Hasinger08}.
The model SEDs of AGN have been tested against the observational relations between X-ray and mid-IR luminosity of AGN recently given  by \citet{Mullaney11}.
The mid-IR LFs predicted from different combinations of the evolution models of HXLF and obscuration of AGN are compared with the measurements of AGN mid-IR LFs given by \cite{Brown06}, \cite{Matute06}, and \cite{Fu10}, respectively.
By predicting mid-IR LFs for type-1 AGN, type-2 AGN, and total AGN from HXLF, and comparing them with corresponding observational results respectively, the evolution of LFs and obscuration of AGN can be further understood.

We find that the mid-IR LFs predicted from HXLFs tend to  underestimate the number of the most IR-luminous AGN, which is independent of the evolution model of AGN obscuration.
We discussed possible explanations for this contradiction.
It may partly due to the missing fraction of Compton-thick AGN that have been predicted by the synthesis model of CXRB, but systematically missed by current X-ray observations.
However, we find that the underestimation to the number of the most IR-luminous  AGN cannot be eliminated even an extreme assumption, which claims that only half of the mid-infrared spectroscopically selected AGN are detected in current X-ray observations, is employed.

We conclude that the contradiction mainly result from limitations in the modeling of the composite X-ray to IR SEDs of AGN.
A possible reason is the contribution of star formation in the AGN host to mid-IR, which has not been considered yet.
We find that the contribution of star formation to the $8.0~\rm{\mu m}$ and $15~\rm{\mu m}$ emission need to be comparable with that of reprocessed nuclear emission and even more in the most IR-Luminous sources to eliminate the contradiction.
However, the contribution of star formation in AGN host to mid-IR is not likely so large but actually decreases with increasing $L_{\rm IR}$.
On the other hand, the contradiction  may represent limitations in the torus model employed.
It is clear that the torus model are further constrained to give the specific prediction of mid-IR LFs for type-1 and type-2 AGN, respectively.
We have made some improvements to the original torus model of \cite{Ballantyne06b}, such as a different handling of the diffuse emission and $L_{\rm UV}$-dependent $R_{\rm in}$, to give the X-ray to mid-IR luminosity ratios that are more comparable to the observational results of \cite{Mullaney11}.
Meanwhile, with some tests we find that the anticorrelation between $\alpha_{\mathrm{ox}}$ and $L_{\rm UV}$ is important for making the X-ray to mid-IR luminosity ratios closer to the results of  \cite{Mullaney11}.
Interestingly, a smaller $\alpha_{\mathrm{ox}}$ improves the prediction of the high-$L_{\rm IR}$ end of the IRLFs significantly at the same time.

Finally, with all the improvements mentioned above, we find that the HXLFs and IRLFs of AGN can be more consistent with each other if the obscuration mechanisms of quasars and Seyferts are assumed to be different.
This is consistent with the idea that the obscuration mechanism of luminous quasars dominating at high redshifts are very different from that of less luminous Seyferts dominating at lower redshifts, corresponding to their different triggering and fueling mechanisms.
However, current measurements of the IRLFs of AGN are not accurate enough to allow a more complete understanding to this by employing the method presented here.
Due to this limitation, the conclusions drawn here need to be tested further when better measurements of IRLFs are available.

More accurate measurements of the IRLFs of AGN, especially those determined at smaller redshift bins and more accurately separated to that for type-1 and type-2, are very helpful for a more complete understanding of the evolution of LFs and obscuration of AGN.
Based on the observations of newly launched IR space telescope such as \spitzer, \herschel~and forthcoming \textit{James Webb Space Telescope (JWST)}, better measurements of the IRLFs of AGN are expected.
These measurements will largely improve our understanding of the evolution of LFs and obscuration of AGN and their co-evolution with galaxies.

\section*{Acknowledgments}
We appreciate the anonymous referee for reviewing our paper very carefully and relevant comments on our work.
We thank the previous anonymous referee for his/her comments and suggestions that largely improved this paper.
This work is supported by the National Natural Science Foundation of China (Grant Nos. 10778702, 11033008, 11063003 and 11103072), the National Basic Research Program of China (Grant No. 2009CB824800) and the Chinese Academy of Sciences (Grant No. KJCX2-YW-T24).

\bibliography{ms.bbl}

\label{lastpage}
\end{document}